\documentclass[aps,pr,twocolumn,
superscriptaddress,groupedaddress,nofootinbib,floatfix]{revtex4}  

\usepackage{soul}
\usepackage{amsfonts}
\usepackage{amssymb}
\usepackage{amsmath}
\usepackage{amsthm}
\usepackage{graphicx}
\usepackage{appendix}
\usepackage{bbold}
\usepackage{dsfont}
\usepackage{enumitem}
\usepackage{xcolor}
\usepackage[
colorlinks=true]{hyperref}

\usepackage{yfonts}
\usepackage{MnSymbol}

\usepackage[normalem]{ulem}
\newcommand\redsout{\bgroup\markoverwith{\textcolor{red}{\rule[0.4ex]{3pt}{0.7pt}}}\ULon}

\providecommand{\U}[1]{\protect\rule{.1in}{.1in}}

\newcommand{\p}{\partial}
\newcommand{\Tr}{\ensuremath{\mathrm{Tr}}}
\newcommand{\tr}{\ensuremath{\mathrm{Tr}}}
\newcommand{\e}{\ensuremath{\mathrm{e}}}

\newcommand{\YM}{\ensuremath{\mathrm{YM}}}
\newcommand{\FP}{\ensuremath{\mathrm{FP}}}
\newcommand{\gf}{\ensuremath{\mathrm{gf}}}
\newcommand{\gh}{\ensuremath{\mathrm{gh}}}

\usepackage{color}
\usepackage{listings}
\definecolor{darkgreen}{rgb}{0,0.35,0}
\definecolor{Rood}{rgb}{1, 0, 0}
\newcommand{\rc}{\textcolor{red}}

\begin{document}

\title{Towards background field independence within the Gribov horizon}



\author{Igor~F.~Justo}
\email{ijusto@id.uff.br}
\affiliation{Instituto de F\'isica, Universidade Federal Fluminense, Campus da Praia Vermelha, Av. Litor\^anea s/n, 24210-346, Niter\'oi, RJ, Brazil}
\author{Antonio~D.~Pereira}
\email{adpjunior@id.uff.br}
\affiliation{Instituto de F\'isica, Universidade Federal Fluminense, Campus da Praia Vermelha, Av. Litor\^anea s/n, 24210-346, Niter\'oi, RJ, Brazil}
\affiliation{Institute for Mathematics, Astrophysics and Particle Physics (IMAPP),
Radboud University, Heyendaalseweg 135, 6525 AJ Nijmegen, The Netherlands}
\author{Rodrigo~F.~Sobreiro}
\email{rodrigo\_sobreiro@id.uff.br}
\affiliation{Instituto de F\'isica, Universidade Federal Fluminense, Campus da Praia Vermelha, Av. Litor\^anea s/n, 24210-346, Niter\'oi, RJ, Brazil}

\begin{abstract}
We introduce a background gauge akin to the Landau-DeWitt gauge but deformed by the presence of a gauge parameter for the quantization of Euclidean Yang-Mills theories. In the limit where the background field vanishes, standard linear covariant gauges are recovered. This gauge allows for an explicit investigation of the effects of infinitesimal Gribov copies and their impact to background and gauge parameter dependence of physical correlators. Similarly to linear covariant gauges, the introduction of gauge-invariant dressed fields is essential to restore BRST symmetry. Hence, we construct a BRST symmetric action in linear covariant background gauges which eliminates regular infinitesimal Gribov copies in analogy to the recently introduced BRST invariant (refined) Gribov-Zwanziger action. The issue of background dependence and its relation to gauge parameter dependence is discussed in the light of non-perturbative effects driven by the elimination of Gribov copies. 
\end{abstract}

\maketitle

\section{Introduction}

Quantum Chromodynamics (QCD) is one of the most fascinating building blocks of the Standard Model of particle physics. At large energy scales, i.e., in the ultraviolet {regime}, the theory exhibits asymptotic freedom \cite{Politzer:1973fx,Gross:1973id}. 
It means that at very short lengths the theory behaves as a \rc{(confined)} free field theory. Towards the infrared, i.e., {at} low energies, the coupling {constant} grows and, perturbatively, hits a Landau pole due to the failure of perturbative methods, since the coupling becomes sufficiently large. Hence, the description of dynamical effects at low energies demands the use of non-perturbative techniques such as lattice simulations, functional methods, effective models {accounting} for topological configurations and non-perturbative effects, and holographic models. We refer the reader to, e.g., \cite{Greensite:2011zz,Brambilla:2014jmp}. Color confinement and chiral symmetry breaking, phenomena that are not described by perturbative tools, are expected to be captured by one or {by} a cross-fertilization of the above mentioned methods. Thence, QCD is a complete theory in the ultraviolet with a very rich dynamics in the infrared and allows one to use and develop highly sophisticated quantum-field theoretic methods for its comprehension. Interestingly, the removal of quarks does not spoil some of the properties just described: gluons are still confined and asymptotic freedom is still present. It means that, potentially, removing quarks can be a first reasonable simplification for the understanding of the mechanism behind confinement due to the already very complicated non-linear dynamics of Yang-Mills theories.

Pure Yang-Mills theories are perturbatively renormalizable and unitary in four dimensions. Observables are gauge invariant and, as in any quantum field theory, they can be constructed out of the (unphysical) gauge-dependent correlation functions. Since we aim at accessing non-perturbative regimes of Yang-Mills theories, the simple computation of correlation functions within perturbation theory is not enough. The computation of gauge-fixed correlation functions in the continuum at strongly-correlated energy scales can be achieved by functional methods such as the functional renormalization group and Schwinger-Dyson equations, see, e.g., \cite{Pawlowski:2003hq,Fischer:2008uz,Dupuis:2020fhh,Cyrol:2016tym,Cyrol:2018xeq,Corell:2018yil,vonSmekal:1997ohs,Alkofer:2000wg,Aguilar:2008xm,Alkofer:2008jy,Binosi:2009qm,Huber:2018ned}. In the present work we take a different avenue where perturbative methods are applied, after taking into account non-perturbative features in the path integral of Yang-Mills theories, though. 
These effects are due to the necessity of improving the Faddeev-Popov gauge-fixing procedure \cite{Faddeev:1967fc} in the infrared. As it is well known since the seminal work by Gribov \cite{Gribov:1977wm} and the rigorous mathematical result by Singer \cite{Singer:1978dk}, there is no local covariant gauge-fixing condition which can completely remove the gauge redundancy and that is continuous in field space. In particular, configurations which satisfy the gauge condition and are connected by gauge transformations are still present in the gauge-fixed path integral of Yang-Mills theories. Such spurious configurations are known as Gribov copies. Removing such configurations from the path integral results in a modification of the Boltzmann weight of gauge-fixed Euclidean Yang-Mills theories, and thereby novel contributions to correlation functions must be taken into account in quantum calculations. Moreover, Gribov copies are not dynamically important in the weakly correlated regime of Yang-Mills theories, leaving the well-established ultraviolet results untouched. Nevertheless, the existence of Gribov copies in the path integral becomes relevant towards the infrared and might be crucial for the determination of the correct low-energy behavior of correlation functions in Euclidean Yang-Mills theories. In particular, in the Landau gauge the removal of infinitesimal Gribov copies together with the effects of non-perturbative generation of dimension-two condensates give rise to the refined Gribov-Zwanziger (RGZ) action \cite{Dudal:2007cw,Dudal:2008sp}, which, aside from being local,  is renormalizable at all orders in perturbation theory
and renders gluon and Faddeev-Popov (FP) ghost propagators at leading order in qualitative agreement with lattice simulations \cite{Cucchieri:2007rg,Maas:2008ri,Bogolubsky:2009dc,Maas:2011se}. The RGZ action is constructed upon the so-called Gribov-Zwanziger (GZ) action which is local, renormalizable, and implements the restriction of the path integral of Euclidean Yang-Mills theories in the Landau gauge to a region free of infinitesimal Gribov copies. This is known as the Gribov region and it features important geometric properties as described in \cite{DellAntonio:1991mms}. The restriction of the functional integral to the Gribov region was worked out at leading order in \cite{Gribov:1977wm} and at all orders using a different prescription in \cite{Zwanziger:1989mf}. The equivalence between those approaches was established in \cite{Capri:2012wx}. Yet the Gribov region is not free of all Gribov copies as pointed out in \cite{vanBaal:1991zw}. For a review of the Gribov problem in the Landau gauge and of the GZ framework, we refer to \cite{Sobreiro:2005ec,Vandersickel:2012tz}. Dynamical infrared instabilities were accounted for in the refinement of the GZ framework leading to the RGZ action in \cite{Dudal:2007cw,Dudal:2008sp}. 

Besides the considerable progress achieved within the RGZ scenario in the Landau gauge, two important conceptual issues were left open until recently: The (R)GZ construction relies on several particular properties of the Landau gauge and in its original formulation, which can be found in  \cite{Zwanziger:1989mf,Dudal:2008sp}, {the} BRST symmetry is explicitly broken. Being a direct outcome of the standard FP quantization, BRST invariance is lost once the path integral is restricted to the Gribov region, i.e., when a modification to the FP gauge-fixing procedure is performed. In the Landau gauge, such a breaking was not taken as an inconsistency of the (R)GZ construction, and it passed by thorough scrutiny \cite{Maggiore:1993wq,Baulieu:2008fy,Dudal:2009xh,Sorella:2009vt,Sorella:2010it,Capri:2010hb,Dudal:2012sb,Capri:2014bsa,Cucchieri:2014via,Schaden:2014bea,Schaden:2015uua}. However, BRST symmetry plays an important role in controlling gauge-parameter dependence of correlation functions. Consequently, dealing with gauge conditions which involve a free gauge parameter poses a challenging task in the removal of Gribov copies if BRST symmetry is generically broken. In \cite{Capri:2015ixa}, it was established how to restore BRST invariance within the RGZ framework by the introduction of a dressed, gauge-invariant field denoted by $A^h_\mu$. By construction, this field collapses to the gauge field itself when Landau gauge is imposed, since the $A^h_\mu$ is essentially {built} as the gauge field {$A_{\mu}$} itself plus {an infinite sum of} non-local contributions {containing the} divergence {$A_{\mu}$} (which vanishes in the Landau gauge). Hence, as pointed out in \cite{Capri:2016gut,Capri:2018ijg}, correlation functions of gauge-invariant operators computed within the standard formulation of the RGZ action in the Landau gauge, where BRST is broken, are equal to those computed in the BRST-invariant formulation of the RGZ action in the Landau gauge. Having a BRST-invariant formulation of the RGZ action in the Landau gauge provides a natural toolbox to extend the action to a generic class of covariant gauges as discussed in \cite{Pereira:2016fpn,Capri:2018ijg}. In particular, this was largely used in the context of linear covariant gauges, \cite{Capri:2015ixa,Capri:2015nzw,Capri:2016aqq,Capri:2016gut,Capri:2017bfd}.

When dealing with finite temperature computations or in the context of applications of functional methods, the use of the background field method (BFM) can play a pivotal role. See, e.g., \cite{DeWitt:1967ub,Weinberg:1996kr,Abbott:1981ke,Binosi:2002ft,Aguilar:2006gr,Canfora:2015yia,Reinosa:2014ooa}. From a broader perspective, the quantization of other theories with gauge symmetries such as gravity might be much better posed by the use of the BFM. However, when thinking in terms of the incompleteness of the FP procedure due to the existence of Gribov copies, the BFM brings new conceptual challenges. A non-dynamical field configuration $\bar{A}_\mu$ is introduced in the gauge-fixing condition. Therefore, it is to be expected that the Gribov copies will depend on the explicit choice of the background field. Hence, eliminating such copies can, eventually, break background independence, i.e., the fact that observables shall not depend on the particular property of the background field. First steps in dealing with the BFM in the presence of Gribov copies were given in \cite{Canfora:2016ngn}. An attempt to follow the original strategy suggested by Gribov was worked out at leading order in \cite{Junqueira:2020whg}. Making use of the toolbox used in the BRST-invariant formulation of the RGZ, a BRST- and background-invariant formulation of the (R)GZ action was proposed in \cite{Dudal:2017jfw}. An alternative proposal was made in \cite{Kroff:2018ncl}. In the present work, we follow the strategy put forward in \cite{Dudal:2017jfw} and construct the BRST-invariant formulation of the RGZ action in a generalized Landau-DeWitt gauge. It corresponds to a gauge condition supporting a background field as well as a gauge parameter. Our main purpose is to establish a clear connection between BRST invariance and background independence. Our findings, although explicit to Yang-Mills theories, might be of relevance to topological field theories, see, e.g., \cite{Astorino:2012af} and continuum quantum-field theoretic formulations of quantum gravity, as the so-called Asymptotic Safety scenario, see \cite{Bonanno:2020bil,Reichert:2020mja}. Background independence is controlled by a Ward identity which encodes the fact that the complete gauge field is written as a sum of a (non-dynamical) background part and quantum fluctuations.

This work is organized as follows: In Sect.~\ref{gpldw}, we introduce our conventions and notation. We define the linear covariant background gauges (LCBG) and introduce the functional identity, the shift Ward Identity (sWI) which controls the background field (in)dependence of the underlying one-particle irreducible generating functional. In Sect.~\ref{0modessec}, the existence of infinitesimal Gribov copies in the LCBG is associated with the zero-modes of the Faddeev-Popov operator. Similarly to standard linear covariant gauges, it is argued that the elimination of zero-modes of the Faddeev-Popov operator restricted to the transverse component of the gauge field is sufficient to eliminate the infinitesimal Gribov copies in the LCBG if they admit a power series in the gauge parameter as well as in the background field. Next, in Sect.~\ref{grestrcsec}, it is proposed the action arising from the elimination of the zero-modes of the transverse projection of the Faddeev-Popov operator. The resulting non-local action is expressed in local form by the introduction of suitable auxiliary fields. The consistency of such a proposal is tested against BRST and background gauge transformations and it is realized that it would be at odds with independence of physical correlation functions from unphysical choices such as the gauge parameter as well as background field. In Sect.~\ref{non-p_BRST} an action which is local, BRST-symmetric, invariant under background gauge transformations, and that effectively restricts the path integral domain to a region free of infinitesimal Gribov copies of a certain class is proposed. It is argued that such an action is consistent with background-field independence of physical correlation functions. Finally, we collect our conclusions.                                    

\section{The setup}
\label{gpldw}

\subsection{The linear covariant background gauges}

In this section we provide all relevant definitions and conventions in the construction of Yang-Mills theory within the BFM in linear covariant gauges. This work is entirely developed in flat Euclidean spacetime, and the corresponding Yang-Mills action describing the dynamics of the gauge field $A_\mu = A^{a}_{\mu}T^{a}$ reads\footnote{In this work we employ the following notation for integrals
\begin{equation}
\int \mathrm{d}^{4}x ~\equiv~ \int_{x}
\,.
\end{equation}
The standard (explicit) notation will be used whenever ambiguity or confusion is possible.} 
\begin{eqnarray}
S_{\mathrm{YM}} ~=~ 
\frac{1}{2}
\tr
\int_{x} F_{\mu\nu} F_{\mu \nu}
\,,
\label{ym1}
\end{eqnarray}
where
$
F_{\mu\nu} =
\p_{\mu}A_{\nu} - \p_{\nu}A_{\mu} - ig[A_{\mu},A_{\nu}] \,\label{fstr1}
$
is the field strength tensor; $g$ stands for the coupling constant; and $T^{a}$ accounts for the $(N^{2}-1)$ generators of the gauge group $SU(N)$. The action \eqref{ym1} is invariant under gauge transformations with $SU(N)$ gauge group. In particular, infinitesimal transformations take the form,
\begin{equation}
\delta A_\mu = -\p_{\mu}\theta + ig[A_{\mu}, \theta] = -D_\mu \theta 
\,,
\label{gt1}
\end{equation}
with $\theta = \theta^{a}T^{a}$ being the infinitesimal gauge parameter, and with the covariant derivative defined as $D_\mu = \p_\mu - ig[A_{\mu}, \ ]$. We will extensively employ the matrix notation. However explicit color index notation will be used whenever it helps in bringing clarity\footnote{
The components in the adjoint representation of $SU(N)$ can be recovered by making explicit the matrix elements of $\{T^{a}\}$,
\begin{align}
	\left[\mathrm{Ad}(A^{a}_{\mu}T^{a})\right]^{bc} = f^{abc}A^{a}_{\mu}
	\,,
\end{align}
along side with the corresponding algebra and the trace relation,
\begin{align}
	[T^{a},T^{b}] = if^{abc}T^{c}
	\quad \text{and} \quad
	\Tr\left( T^{a}T^{b} \right) = \frac12 \delta^{ab}
	\label{grpgen0}
	\,,
\end{align}
with $f^{abc}$ representing the real and totally anti-symmetric structure constants of $SU(N)$.}.

The BFM relies on the introduction of a non-dynamical gauge field $\bar{A}_{\mu}$ in such a way that the full gauge field $A_\mu$ is written as $A_\mu = \bar{A}_\mu + a_\mu$, with $a_\mu$ denoting dynamical fluctuations about the background $\bar{A}_\mu$. The original YM action \eqref{ym1} becomes $S_{\YM}[A_{\mu}] = S_{\YM}[\bar{A}_{\mu} + a_{\mu}]$. Clearly, choosing a trivial background ($\bar{A}_\mu = 0$) {it falls} 
back {to} the standard form of the YM action. The explicit form of the YM action within the BFM reads \cite{Weinberg:1996kr}, 
\begin{align}
S_{\YM} = 
\frac{1}{2}\int_{x}\tr\,\Big( \bar{F}_{\mu\nu} + \bar{D}_\mu a_\nu -
\bar{D}_\nu a_\mu - ig\,[a_{\mu},a_{\nu}] \Big)^{2}
\,,
\label{ym2}
\end{align}
where
$
\bar{D}_{\mu} = \p_{\mu} - ig\,[\bar{A}_{\mu},\ ]
\,,
$
and 
$\bar{F}_{\mu\nu} =
\p_\mu\bar{A}_\nu - \p_\nu\bar{A}_{\mu} - ig [\bar{A}_\mu , \bar{A}_\nu]$.

As it is well known, the presence of $\bar{A}_{\mu}$ does not spoil gauge symmetry, \cite{DeWitt:1967ub,Abbott:1981ke,Weinberg:1996kr}. In fact, the new action \eqref{ym2} is invariant under the following infinitesimal gauge transformations,
\begin{align}
    \delta \bar{A}_\mu = 0 \,,
    \qquad
    \delta a_\mu = -D_{\mu} \theta
\,,
\label{gt2}
\end{align}
where $
D_{\mu}\theta = \p_{\mu}\theta - ig [\bar{A}_{\mu} + a_{\mu}, \theta]$ is the ``complete" covariant derivative. Therefore, the Faddeev-Popov gauge fixing procedure can be applied, and we impose a linear covariant extension of the Landau-DeWitt (LDW) gauge, see, e.g, \cite{Binosi:2013cea} -- hereby named as linear covariant background gauge (LCBG),
\begin{equation}
\bar{D}_{\mu} a_{\mu} - \alpha b = 0 
\,,
\label{lcgdwgaugecond} 
\end{equation}
with $\alpha$ being a non-negative gauge parameter. At this point the LDW gauge condition is recovered by taking the limit $\alpha \to 0$. Likewise, standard linear covariant gauges are obtained by taking $\bar{A}_\mu = 0$, with Landau gauge being a particular case with $\alpha \to 0$. By means of the Faddeev-Popov gauge fixing procedure one gets the following partition function
\begin{equation}
    {\cal Z}[\bar{A}]=
    \int[{\rm d}a][{\rm d}\bar{c}][{\rm d}c][{\rm d}b]\,
    \e^{-S_{\FP}}
    \,,
    \label{z1}
\end{equation}
with $(\bar{c},c)$ denoting the Faddeev-Popov ghosts and $b$ the Lautrup-Nakanishi auxiliary field. The Faddeev-Popov action reads,
\begin{equation}
	{S}_{\FP} = 
	{S}_{\YM} +
	{S}_{\mathrm{gf}} + 
	{S}_{\mathrm{gh}}
	\,.
	\label{fpaction0}
\end{equation}
The Yang-Mills action ${S}_{\YM}$ is given by \eqref{ym2}; the sum of the gauge fixing and Faddeev-Popov ghost actions yields
\begin{equation}
    {S}_{\mathrm{gf}} + 
    {S}_{\mathrm{gh}} =
    \int_{x}
    \left\{
     b^{a} 
     \left(
     \bar{D}^{ab}_{\mu}a^{b}_{\mu} - \frac{\alpha}{2}b^{a}
     \right)
     + 
     \bar{c}^{a}
     {\mathcal{M}}^{ab} c^{b}
     \right\}
    \,,
    \label{gf1}
\end{equation}
and the Faddeev-Popov operator is given by 
\begin{align}
{{\cal M}}^{ab}(a_{\mu},\bar{A}_{\mu}) 
&=-\bar{D}^{ad}_\mu D^{db}_\mu 
\,.
\label{fpop}
\end{align}
Here ${\cal M}$ is not Hermitian, unlike to the standard LDW gauge since with the LCBG condition \eqref{lcgdwgaugecond} the fluctuation gauge field $a_{\mu}$ gets a non-vanishing longitudinal component. Moreover, the operator ${\cal M}$ falls back into the usual Faddeev-Popov operator in linear covariant gauges by turning off the background field, $\mathcal{M}^{ab}(a_{\mu},0) = \mathcal{M}^{ab}(a_{\mu}) = -\p_{\mu}D_{\mu}$, which is not Hermitian as well. Hermiticity plays an important role for the elimination of Gribov copies since the infinitesimal ones are associated with zero-modes of the Faddeev-Popov operator. When the operator is Hermitian, it is possible to define the set of gauge field configurations for which the Faddeev-Popov operator is positive due to the real nature of its spectrum. As such, one is able to define a region free of infinitesimal copies where $\mathcal{M}$ assumes only positive definite eigenvalues.

\subsection{Shift symmetry}

At this stage is also worth mentioning what are the effects of the gauge-fixing term on the background field structure. In the pure Yang-Mills action, the background gauge field $\bar{A}_\mu$ is always accompanied by the fluctuation field $a_\mu$ in a sum, i.e., $S_{\YM}[A] = S_{\YM} [\bar{A}+a]$, reflecting that the theory depends on a single field $A_\mu$. We can then define a shift transformation as,
\begin{equation}
    \bar{A}^{\prime}_\mu = \bar{A}_\mu - \epsilon_\mu (x)\,,\qquad a^\prime_\mu = a_\mu + \epsilon_\mu (x)\,,
    \label{shiftsymmetry1}
\end{equation}
leading to the following symmetry, 
\begin{equation}
    S_{\YM}[\bar{A}^\prime+a^\prime] = S_{\YM}[\bar{A}+a] = S_{\YM}[A]\,.
    \label{shiftsymmetry2}
\end{equation}
This can be converted into a classical functional identity as
\begin{equation}
    \frac{\delta S_{\YM}}{\delta a_\mu} - \frac{\delta S_{\YM}}{\delta \bar{A}_\mu} = 0\,.
    \label{ClassicalShiftWI}
\end{equation}
However, the introduction of the gauge-fixing action \eqref{gf1} breaks the appearance of background and fluctuation fields just as a sum. Hence, the shift-symmetry identity will, in principle, be broken by contributions arising from the gauge-fixing term.

Let us consider the partition function \eqref{z1} in the presence of an external source $J^a_\mu$ coupled to the fluctuation $a^a_\mu$ as follows,
\begin{equation}
    {\cal Z}[J,\bar{A}] = \int [{\rm d}a][{\rm d}\bar{c}][{\rm d}c][{\rm d}b]\,\e^{-S_{\FP}+\int_x J^{a}_\mu a^a_\mu}\,.
\end{equation}
Performing an infinitesimal shift transformation over $\bar{A}_\mu$, i.e., $\bar{A}_\mu \to \bar{A}^\prime_\mu = \bar{A}_\mu-\epsilon_\mu (x)$, and renaming the dummy variable $a^a_\mu$ to $a^{\prime\, a}_\mu$ in such a way that $a^{\prime\, a}_\mu = a^{a}_\mu+\epsilon^a_\mu$, one can derive the following identity,
\begin{eqnarray}
    {\cal B}(\hat{a},\bar{A})\circ \Gamma = \langle {\cal B}(a,\bar{A})\circ (S_{\gf}+S_{\gh})\rangle_{J,\bar{A}}\,,
\label{ShiftWI1}
\end{eqnarray}
where $\Gamma$ stands for the one-particle irreducible generating functional and with
\begin{equation}
    {\cal B}(\varphi,\bar{\phi})\circ F = \frac{\delta F}{\delta \varphi(x)} - \frac{\delta F}{\delta \bar{\phi}(x)}\,.
\label{ShiftWI2}
\end{equation}
Identity \eqref{ShiftWI1} is called shift or split Ward identity. Essentially, it establishes a relation between derivatives with respect to the background field and derivatives with respect to the fluctuations. It is a  consequence of the fact that the full theory describes the dynamics of a single field, i.e., $A_\mu$. The field $\hat{a}^a_\mu$ represents the expectation value of $a^a_\mu$ at non-vanishing source. The derivation of \eqref{ShiftWI1} is standard and is reported on Appendix~\ref{Ap:ShiftWI} for completeness. It is also clear from eq.\eqref{ShiftWI1} that the quantum extension of \eqref{ClassicalShiftWI} is ``broken" by the necessity of introducing {the} gauge-fixing and FP ghost terms. Yet such a ``breaking" is harmless due to the fact that the gauge-fixing procedure leads to a BRST-exact contribution. 
For instance, according to our particular gauge choice \eqref{gf1}, one is able to obtain 
\begin{equation}
    \langle \mathcal{B}^a_\mu (a,\bar{A})\circ (S_{\rm gf}+S_{\rm gh})\rangle = - \Big\langle s\Big(D^{ab}_\mu\bar{c}^b\Big)\Big\rangle_{\bar{A},J}\,.
    \label{PartitionFunction9}
\end{equation}
This establishes a clear relation between background field independence (i.e. the shift symmetry) and BRST symmetry, meaning that an explicit breaking of BRST symmetry would put in risk the background field independence of physical quantities. We shall return to those issues in Sect.~\ref{non-p_BRST}.

\section{The zero-modes of the Faddeev-Popov operator}
\label{0modessec}

Substantial progress was made over the past decades in removing {infinitesimal} Gribov copies from the gauge-fixed functional integral of Yang-Mills theories. {These} copies are generically connected to the existence of normalizable zero-modes of the Faddeev-Popov operator in the corresponding chosen gauge. Hence, one can track {these} infinitesimal Gribov copies by investigating the zero-modes of the Faddeev-Popov operator. This section is devoted to this issue in the LCBG.

Let us start by defining the infinitesimal gauge equivalent configurations (or gauge copies): ${a}^\prime_{\mu}$ and $a_{\mu}$
are said to be gauge equivalent configurations if both of them satisfy the LCBG condition \eqref{lcgdwgaugecond} and are linked to each other by an infinitesimal gauge transformation,
\begin{align}
\bar{D}_{\mu}{a}^\prime_{\mu} &= \bar{D}_{\mu}a_{\mu} = \alpha b\;,
\nonumber \\
a^\prime_{\mu} &= a_{\mu} - D_{\mu} \theta\;.
\label{gc1}
\end{align}
Hence, infinitesimal gauge copies exist if the infinitesimal gauge parameter $\theta$ is a zero-mode of the Faddeev-Popov operator, i.e.,
\begin{align}
{\mathcal{M}}^{ab}(a_\mu,\bar{A}_\mu)\theta^b\equiv -\bar{D}^{ad}_\mu D^{db}_\mu \theta^b = 0
\,.\label{gc2}
\end{align}
The existence of normalizable zero-modes is the central issue of the Gribov problem at the level of infinitesimal copies. Actually, given their existence, the biggest challenge is to get rid of them from the path integral measure. In the Landau(-DeWitt) gauge such a task is simplified due to the Hermiticity of the Faddeev-Popov operator, which allows us to define a region of configuration related to positive definite eigenvalues of $\mathcal{M}$. As mentioned before, in linear covariant gauges (within or without the BFM) the Faddeev-Popov operator is not Hermitian, which forbid us from using the usual Gribov method in the Landau gauge. An interesting strategy was developed in \cite{Sobreiro:2005vn,Capri:2015pja} for linear covariant gauges.

In the present section we repeat the same strategy of \cite{Capri:2015pja} {adapted} to the LCBG, which we expect to lead to a BRST breaking, just as in \cite{Capri:2015pja}. The compatibility between BRST symmetry and the restriction process will be developed in Section \ref{non-p_BRST}. After all, standard linear covariant gauges are a particular case of LCBG with the background chosen as the trivial one, i.e., $\bar{A}_\mu = 0$. Yet repeating such an exercise is helpful as an intermediate step for the construction of the BRST-invariant formulation and we keep it for the benefit of the reader.

The proposal is to decompose the total field $A_{\mu}$ into longitudinal and transverse components,
\begin{align}
A_{\mu} &= A^{T}_{\mu} + A^{L}_{\mu}
\,,
\label{decomp.TL}
\end{align}
with 
\begin{equation}
A^{T}_{\mu} = a^{T}_{\mu}  + \bar{A}_{\mu}
\equiv \left( \delta_{\mu\nu} - \frac{\p_{\mu}\p_{\nu}}{\p^{2}} \right) A_{\nu}\,, 
\label{transgauge}
\end{equation}
and
\begin{align}
A^{L}_{\mu} &= a^{L}_{\mu}
\equiv  
\frac{\p_{\mu}\p_{\nu}}{\p^{2}}   A_{\nu} 
\,.
\label{longit}
\end{align}
For the sake of simplicity and clarity of notation, from now on the background and fluctuation configurations $(\bar{A}_{\mu}+a_{\mu})$ will be kept hidden into the total configuration $A_{\mu}$, whenever there is no ambiguity, otherwise they will be made explicit. Furthermore, notice that we are assuming only transverse configurations for the background field, so that it only enters in the transverse component of $A_{\mu}$.

From the gauge fixing condition \eqref{lcgdwgaugecond} one is able to express $A^{L}_{\mu}$ as,
\begin{align}
A^{L}_{\mu} = \frac{\p_{\mu}}{\p^{2}} \left( \alpha b + ig [\bar{A}_{\nu}, A_{\nu}] \right)
\,.
\label{Al0}
\end{align}
This is done in order to establish a result closely related to the one in \cite{Sobreiro:2005vn,Capri:2015pja}: by imposing positivity to the Faddeev-Popov like operator restricted only to transverse configurations\footnote{When restricted to transverse configurations, the FP operator is Hermitian.} $A^{T}_{\mu}$ is enough to get rid of zero-modes (expandable in powers of $\alpha$ and of the background field) of the actual Faddeev-Popov operator \eqref{fpop}:
\begin{itemize}
\item
\textit{\textbf{Statement:}} 
If $A^{T}_{\mu}$ is such
that the operator
\begin{align}
\mathcal{M}^{T}(a^{T},\bar{A}) &=
- \p_{\mu}\left( \p_{\mu} - ig [A^{T}_{\mu}, \ ] \right)
\nonumber \\
&\equiv
-\p_{\mu}D^{T}_{\mu}\,,
\,
\label{Cop}
\end{align}
is positive definite, then the Faddeev-Popov operator \eqref{fpop} does not develop zero-modes for $A_\mu = A^{T}_\mu + A^L_\mu$.

\item
\textit{\textbf{Proof:}} 
First, notice that ${\mathcal{M}}^{T}$ is Hermitian, hence the positivity assumption makes sense. Now, let us consider the zero-mode equation for the Faddeev-Popov operator,
\begin{align}
-\bar{D}_{\mu}D_{\mu}\theta = 0
\;.
\label{fp0modeeq}
\end{align}
By decomposing $A_{\mu}$ into $A^{T}_{\mu} + A^{L}_{\mu}$ we get
\begin{align}
{\mathcal{M}}^{T}\theta  = - ig \left([\bar{A}_{\mu}, D_{\mu}\theta] + \p_{\mu}[A^{L}_{\mu},\theta] \right)
\,.
\end{align}
Since $A^{T}_{\mu}$ ($=a^{T}_{\mu} + \bar{A}_{\mu}$) is such that ${\mathcal{M}}^{T}$ is positive definite, then the zero-mode equation can be rewritten as
\begin{align}
&
\theta = 
-ig \left[{\mathcal{M}}^{T}\right]^{-1} 
\left(
\p_{\mu}[A^{L}_{\mu}, \theta] + [\bar{A}_{\mu}, D_{\mu}\theta]
\right)
\,.
\end{align}
From \eqref{Al0} we have
\begin{align}
\theta  &= 
-ig \left[{\mathcal{M}}^{T}\right]^{-1} 
\bigg(
\alpha\p_{\mu}\left[\frac{\p_{\mu}}{\p^{2}}b, \theta \right] + 
\nonumber \\
&
+ 
ig\p_{\mu} \left[\frac{\p_{\mu}}{\p^{2}}[\bar{A}_{\nu}, A_{\nu}], \theta \right] +
[\bar{A}_{\mu}, D_{\mu}\theta]
\bigg)
\,.
\label{eqxi}
\end{align}
At this point let us assume that $\theta$ is a smooth function of $\alpha$ and $\bar{A}_{\mu}$, in the sense that it can be expanded as
\begin{equation}
\theta = \sum_{l,n=0}^{\infty} \theta_{l,n}\, \alpha^{l}(\bar{A}_{\mu}\bar{A}_{\mu})^{n}
\label{thetaexpanded}
\,,
\end{equation}
so that the general coefficient $\theta_{l,n}$ corresponds to the term of $l$-th power of $\alpha$ and $n$-th power of $\bar{A}_{\mu}\bar{A}_{\mu}$. Hence, by replacing \eqref{thetaexpanded} into \eqref{eqxi}, one is able to verify, recursively, that $\theta_{l,n} = 0$.

\end{itemize}

Consequently, it suffices to impose the restriction of the path integral over $a^{T}_{\mu}$ to the region
\begin{equation}
{\Omega}^{T} = 
\left\{
A^{T}_{\mu} \,;~ \bar{D}_{\mu} A_{\mu} = \alpha b, ~{\mathcal{M}}^{T}(a^{T},\bar{A}) >0
\right\}
\,,
\label{greg01}
\end{equation}
in order to get the non-Hermitian Faddeev-Popov operator ${\mathcal{M}}$ rid of infinitesimal zero-modes. Notice that despite the region ${\Omega}^{T}$ is defined in terms of $A^{T}_{\mu} = a^{T}_{\mu} + \bar{A}_{\mu}$, the functional measure of the path integral is not affected, in the sense that $\bar{A}_{\mu}$ is a classical configuration and $[\mathrm{d}A^{T}] = [\mathrm{d}a^{T}]$. The region ${\Omega}^{T}$ plays a similar role to the Gribov region in the Landau gauge. Clearly, in a trivial background $\bar{A}_\mu = 0$ and with $\alpha = 0$, ${\Omega}^{T}$ recovers the Gribov region $\Omega$.

\section{Implementing the restriction of the path integral domain}
\label{grestrcsec}

Once the region free of infinitesimal Gribov copies is determined, the functional integral of the gauge field must be restricted to that domain for the elimination of such spurious configurations. As we have seen in the last section, in the LCBG it is enough to impose such a restriction to the transverse component of $A_{\mu}$.

Following the standard procedure to impose such a restriction in scenarios without the BFM, either in the
Landau gauge,
\cite{Zwanziger:1988jt,Sorella:2010fs,Vandersickel:2012tz,Dudal:2012sb},
or in linear covariant gauges,
\cite{Sobreiro:2005vn,Capri:2015pja,Capri:2015ixa,Capri:2016aqq}, the
restricted path integral to ${\Omega}^{T}$ shall be given by\footnote{{In the following expression we have omitted the integration over the Faddeev-Popov ghosts and the Lautrup-Nakanishi field for simplicity.}}
\begin{align}
\int_{{\Omega}^{T}} [\mathrm{d}a^{T}][\mathrm{d}a^{L}]\, \e^{-S_{\FP}} = \int [\mathrm{d}a^{T}][\mathrm{d}a^{L}]\,
\e^{-{S}}
\,,
\label{wghi41}
\end{align}
with
\begin{align}
{S} = {S}_{\FP} + {S}_{H^{T}} - \gamma^{4}d(N^{2}-1)
\label{nlgzaction0}
\end{align}
and
\begin{align}
{S}_{{H}^{T}} &= 
\gamma^{4}g^{2}
\int_{x} 
f^{abc}
{A^{T}}^{a}_{\mu} 
\left[ \left({\mathcal{M}}^{T}\right)^{-1}  \right]^{ce}
f^{dbe}
{A^{T}}^{d}_{\mu}
\nonumber \\
&=
\gamma^{4} {H}^{T}(a^{T},\bar{A})
\,,
\label{nlhf}
\end{align}
where the ${H}^{T}(a^{T},\bar{A})$ stands for the Horizon function within the BFM in the LCBG, and $\gamma$ is the so-called Gribov parameter. Perturbatively, since $A^{T}_{\mu} = a^{T}_{\mu} + \bar{A}_{\mu}$, one must expect contributions to the Horizon function that depends only on the background $\bar{A}_{\mu}$ and on linear terms of $a^{T}_{\mu}$. In the LDW gauge these terms were verified at leading order by following the original Gribov no-pole condition, \cite{Junqueira:2020whg}.

Due to the presence of $\left({\mathcal{M}}^{T}\right)^{-1}$ and to the definition of $A^{T}_{\mu}$, given by \eqref{transgauge}, the Horizon function has two sources of non-localities, which can be made local by means of the procedure detailed in \cite{Capri:2015pja}. 

Let us first pay attention to the non-locality of $A^{T}_{\mu}$, which can be worked around by rewriting this component in terms of $a_{\mu}$ and of a new auxiliary field $h^{a}_{\mu}$,
\begin{equation}
A^{T}_{\mu} = A_{\mu} - h_{\mu}
\,.
\end{equation}
Notice that once we have replaced $A^{T}_{\mu}$ by $A_{\mu} - h_{\mu}$
in \eqref{nlhf}, two conditions must be satisfied,
\begin{equation}
h_{\mu} = A^{L}_{\mu}
\quad \text{and} \quad
{\partial_\mu} (A_{\mu} - h_{\mu}) = 0
\,.
\label{cond0}
\end{equation}
To impose {both} conditions \eqref{cond0}, let us introduce two contributions to the action \eqref{nlhf}:
\begin{align}
{S}_{\Lambda\lambda} &= 
\int_{x} 
\lambda^{a}_{\mu} \left( \p^{2}h^{a}_{\mu} 
- \p_{\mu}\p_{\nu}A^{a}_{\nu} \right)\nonumber \\
&-
\int_{x} \Lambda^{a}_{\mu} \left( \p^{2} \xi^{a}_{\mu} -
\p_{\mu}\p_{\nu} D^{ab}_{\nu} c^{b} \right)
\,,
\label{brstinvcond1}
\end{align}
and
\begin{align}
{S}_{\tau v} =
\int_{x} v^{a} \p_{\mu} (A^{a}_{\mu} - h^{a}_{\mu}) +
\int_{x} \tau^{a} \p_{\mu} \left( D^{ab}_{\mu}c^{b} + \xi^{a}_{\mu}
\right)
\,,
\label{brstinvcond2}
\end{align}
where $\lambda_{\mu}$ and $v$ play the role of Lagrange multipliers to
impose, respectively, the first and second conditions of \eqref{cond0}. The auxiliary fields $(h_{\mu}, \xi_{\mu})$, $(\lambda_{\mu}, \Lambda_{\mu})$ and $(v, \tau)$ were conveniently introduced as BRST doublets ({\it cf.} equation \eqref{brsttransf1}).

In the sequence, the non-locality that steams from $\left[{\mathcal{M}}^{T}\right]^{-1}$ can be worked around by the introduction of a set of auxiliary fields
$\left\{ \bar{\omega}_{i}, \omega_{i}, \bar{\varphi}_{i},\varphi_{i} \right\}$, such that once they are integrated out the
non-local version \eqref{nlhf} is recovered. The $i$-index in these auxiliary fields stands for the multi-index notation, which accounts for the pair of spacetime Lorentz and color index: $i = \{\mu, b\}$. 

The localized Horizon function reads,
\begin{align}
\gamma^{4}{H}^{T}_{\text{loc}} &=
\int_{x} 
\Big\{
\bar{\omega}^{a}_{i} 
\left({\mathcal{M}}^{T}\right)^{ab}
\omega^{b}_{i}
-\bar{\varphi}^{a}_{i} 
\left({\mathcal{M}}^{T}\right)^{ab}
\varphi^{b}_{i}-
\nonumber \\
&
- gf^{abc}
\p_{\nu} \bar{\omega}^{a}_{i} 
{D^{T}}^{cd}_{\nu}c^{d} \varphi^{b}_{i}
- gf^{abc}
\p_{\nu} \bar{\omega}^{a}_{i} \xi^{c}_{\nu} \varphi^{b}_{i}
\Big\} +
\nonumber \\
&+
\gamma^{2}g \int_{x}
f^{abc}
\left(A^{a}_{\mu} - h^{a}_{\mu}\right)(\bar{\varphi}^{bc}_{\mu} + \varphi^{bc}_{\mu})
\,.
\label{locgz1}
\end{align}

Finally, the local version of \eqref{nlgzaction0} is given by the sum of the {contributions} \eqref{fpaction0}, \eqref{brstinvcond1}, \eqref{brstinvcond2} and  \eqref{locgz1}:
\begin{equation}
{S}_{\mathrm{Loc}} = {S}_{\FP} +
 {S}_{\Lambda\lambda} + {S}_{\tau v} + \gamma^{4}{H}^{T}_{\text{loc}} - \gamma^{4}d(N^{2}-1)
\,.
\label{locact1}
\end{equation}

The BFM was developed as a valuable tool for keeping track of an auxiliary symmetry which acts as a sort of gauge transformation of the background field, \cite{DeWitt:1967ub,Abbott:1981ke,Weinberg:1996kr}. As {it} will be reviewed here, such a background gauge ($B$-gauge) transformation still is a symmetry of the Faddeev-Popov action \eqref{fpaction0} in the LCBG. However, the main point of this work is to discuss the status of the BRST and $B$-gauge symmetries after performing the Gribov restriction, \emph{i.e.} of the action \eqref{locact1} as just outlined.

\subsection{The BRST symmetry}

As it is well known, the Faddeev-Popov action \eqref{fpaction0} is invariant under BRST transformations, see, e.g., \cite{Baulieu:1981sb}. The fields that appear in ${S}_{\FP}$ transform as
\begin{align}
s\bar{A}_{\mu} &= 0\,;
\quad
sa_{\mu} = -D_{\mu}c
\,;
\label{brsta}
\\
s\bar{c} &= b\,;
\quad 
s b = 0\,;
\label{brstcbarb}
\\
sc &= -\frac12 ig[c,c]
\,.
\label{brsttransf0}
\end{align}
As mentioned before, the auxiliary fields were introduced as BRST doublets, i.e.,
\begin{align}
&
sh_{\mu} = \xi_{\mu}\,, \quad 
s\xi_{\mu} = 0 \,, \quad 
\nonumber \\
&
s\Lambda_{\mu} = \lambda_{\mu} \,, \quad
s\lambda_{\mu} =0 \,, \quad
\label{brsttransf1}
\\
&
s\tau = v \,, \quad
sv = 0
\,,
\nonumber 
\end{align}
so that the contributions $\bar{S}_{\Lambda\lambda}$ and $\bar{S}_{\tau v}$ can be rewritten as BRST-exact terms:
\begin{equation}
{S}_{\Lambda\lambda} = s\int 
\Lambda^{a}_{\mu}\left[ \p^{2}h^{a}_{\mu} 
- \p_{\mu}\p_{\nu}A^{a}_{\nu}
\right]
\,;
\label{brstinvcond12}
\end{equation}
and
\begin{equation}
{S}_{\tau v} =
s\int_{x} 
\tau^{a} \p_{\mu} (A^{a}_{\mu} - h^{a}_{\mu})
\,.
\label{brstinvcond22}
\end{equation}
Hence, they belong to the trivial part of the cohomology of the BRST operator.

The most interesting content of the action \eqref{locact1} comes from
$\gamma^{4}{H}^{T}_{\text{loc}}$. The auxiliary fields $\left\{ \bar{\omega}_{i}, \omega_{i}, \bar{\varphi}_{i}, \varphi_{i} \right\}$ does also transform as doublets of the BRST operator,
\begin{align}
s\bar{\omega}_{i} &= \bar{\varphi}_{i} \,, \quad
s\bar{\varphi}_{i} = 0\,,
\\
s\varphi_{i} &= \omega_{i} \,, \quad
\omega_{i} = 0\,,
\end{align}
{and} as a consequence, the Horizon function \eqref{locgz1} can be written as
\begin{align}
\gamma^{4}{H}^{T}_{\text{loc}} &= 
- s\int_{x}
\bar{\omega}^{a}_{i} 
\left({\mathcal{M}}^{T}\right)^{ab}
\varphi^{b}_{i} 
+
\Delta_{\gamma^{2}} 
\,,
\label{bstgz0}
\end{align}
with 
\begin{align}
\Delta_{\gamma^{2}} = 
\gamma^{2} g\int_{x} 
f^{abc}
\left(A^{a}_{\mu} - h^{a}_{\mu}\right)(\bar{\varphi}^{bc}_{\mu} + \varphi^{bc}_{\mu})
\,.
\label{gammaterm}
\end{align}
Notice that $\Delta_{\gamma^{2}}$ is \emph{not} BRST invariant. Actually, the BRST variation of $\Delta_{\gamma^{2}}$ reads
\begin{align}
s\Delta_{\gamma^{2}}
&=
- \gamma^{2} g f^{abc}
\int_{x} 
\Big\{ 
\left[ D_{\mu} c + \xi_{\mu} \right]^{a}
\left( \bar{\varphi}^{bc}_{\mu} + \varphi^{bc}_{\mu} \right) 
-
\nonumber \\
&
\phantom{\gamma^{2} g f^{abc}\int_{x} \Big\{ }
-
\left( A^{a}_{\mu} - h^{a}_{\mu} \right) \omega^{bc}_{\mu}
\Big\} \neq 0
\,,
\label{brstbreaking}
\end{align}
which means that the BRST symmetry is (softly) broken due to the restriction to ${\Omega}^T$, just as in the scenario without {$\bar{A}_{\mu}$}, see, \cite{Capri:2015pja}. Notice that if the Gribov parameter $\gamma^2$ is set to zero, then the BRST-breaking term vanishes. Since BRST is not a symmetry of the proposed action that eliminates infinitesimal Gribov copies, according to the discussion about the sWI in Sect.~\ref{gpldw} and in Appendix~\ref{Ap:ShiftWI}, it is to be expected that background field independence will not hold. In the sequence we will see that, besides the BRST, the Gribov term $\Delta_{\gamma^{2}}$ does also break the $B$-gauge symmetry.

\subsection{The $B$-gauge transformation}

A well-known fact is that the Faddeev-Popov action is invariant under a background gauge transformation where the fields transform as, \cite{DeWitt:1967ub,Abbott:1981ke,Weinberg:1996kr},
\begin{align}
	\delta_{B}\bar{A}_{\mu} &= -\bar{D}_{\mu}\theta\,,
	\nonumber \\
	\delta_{B} \phi &= -ig[\phi,\theta] \,,
	\nonumber \\
	\delta_{B}\bar{c} &= -ig[\bar{c},\theta] \,,
	\nonumber \\
	\delta_{B}{c} &= -ig[{c},\theta] \,.
	\label{bgaugerule0}
\end{align}
Under the $\delta_{B}$ variation the auxiliary fields $\left\{\Lambda^{a}_{\mu}, \lambda^{a}_{\mu}\right\}$, $\left\{\tau^{a}, v^{a} \right\}$ and $\left\{\bar{\omega}^{ab}_{\mu},\omega^{ab}_{\mu}, \bar{\varphi}^{ab}_{\mu}, \varphi^{ab}_{\mu}\right\}$ transform as

\begin{align}
	\delta_{B}\Lambda_{\mu} &= -ig[\Lambda_{\mu},\theta] \,,
	\nonumber \\
	\delta_{B}\lambda_{\mu} &= -ig[\lambda_{\mu},\theta] \,,
	\nonumber \\
	\delta_{B}v &= -ig[v,\theta] \,,
	\label{bgaugerule2}
	\\
	\delta_{B}\tau &= -ig[\tau,\theta] \,,
	\nonumber \\
	\delta_{B} \phi^{ab} &= -ig{\phi}^{mb}_{\mu} T^{b}[T^{m},\theta^{e}T^{e}]  -ig{\phi}^{am}_{\mu}T^{a}[T^{m},\theta^{e}T^{e}] \,,
	\nonumber 
\end{align}

with $\phi^{ab}$ collectively denoting the auxiliary fields $\left\{\bar{\omega}^{ab}_{\mu},\omega^{ab}_{\mu}, \bar{\varphi}^{ab}_{\mu}, \varphi^{ab}_{\mu}\right\}$.

Before taking the variation of \eqref{locact1} under $\delta_{B}$, let us verify that the $B$-gauge transformation commutes with BRST.

Let $\phi=\phi^{a}T^{a}$ stands for every field carrying a color index (a similar procedure can be developed for fields with two color indices). Besides, assume that $\tilde{\phi} = s\phi$. Thus, the BRST transformation of $\delta_{B}\phi^{a}$ reads,
	\begin{align}
		s(\delta_{B}\phi) &= -ig\,s[\phi,\theta]
		\nonumber \\
		&=
		-ig[s\phi,\theta]
		\nonumber \\
		&= 
		-ig[\tilde{\phi},\theta]
		\nonumber \\
		&=
		\delta_{B}(s\phi)
		\,.
		\label{bBRSTc0}
	\end{align}
Above we have used the fact that $s\theta = 0$.

Now let us proceed to the variation of the local action \eqref{locact1} under $\delta_{B}$. This is a rather easy task, since the contributions from ${S}_{\Lambda\lambda}$ and ${S}_{\tau v}$ are BRST-exact terms, and thus $\delta_{B}{S}_{\Lambda\lambda}$ and $\delta_{B}{S}_{\tau v}$ can schematically be written as 
\begin{align}
\delta_{B}\, s\int_{x} \phi_{1}^{a}\phi_{2}^{a} = s\, \delta_{B}\int_{x} \phi_{1}^{a}\phi_{2}^{a}
\,,    
\end{align}
which is trivially $B$-gauge invariant.

The last and most interesting sector is the Horizon function, $\gamma^{4}{H}^{T}_{\text{loc}}$. As written in eq.\eqref{bstgz0}, the Horizon function can be written as the sum of two terms: one that is trivially BRST invariant; and other one that (softly) breaks the BRST symmetry. For the trivially BRST invariant term we have, 
\begin{align}
\delta_{B} \int_{x} 
s\left[
\bar{\omega}^{ad}_{\mu} 
\left({\mathcal{M}}^{T}\right)^{ab} \varphi^{bd}_{\mu} 
\right]
= s \int_{x} \delta_{B}(\phi^{ad}_{1}\phi^{ad}_{2})  
= 0
\,.
\end{align}
For the $\Delta_{\gamma^{2}}$ term we have,
\begin{align}
\delta_{B}\Delta_{\gamma^{2}} &= 
\gamma^{2} g f^{abc}
\int_{x} 
\Big\{ \left[ -D_{\mu}^{ad}\theta^{d} {- gf^{ade}h^{e}\theta^{d}} \right](\bar{\varphi}^{bc}_{\mu} + \varphi^{bc}_{\mu})
\nonumber \\
&\neq 0
\,.
\label{bgaugegamma}
\end{align}

It is interesting to notice that the term that breaks BRST is the same that breaks the $B$-gauge symmetry, and is the only one that carries the Gribov parameter $\gamma^{2}$.

It is clear that breaking BRST as well as $B$-gauge symmetries will lead to worrisome consequences to the underlying quantum theory. At a technical level, breaking $B$-gauge invariance will lead to the proliferation of terms containing the background field $\bar{A}_\mu$ which are not gauge covariant. This is not only a severe drawback in the use of the BFM but also would require a very cautious treatment of the possible counterterms that are generated in such a theory since there would be no reason to forbid the generation of several of them that do not respect $B$-gauge invariance. Clearly, the breaking of $B$-gauge symmetry is intimately related to the soft breaking of BRST symmetry, cf. \cite{Dudal:2017jfw}. Breaking BRST symmetry also brings, at least, two sources for spurious dependencies in correlation functions of gauge-invariant operators, i.e., observables. It is BRST symmetry that ensures the independence of observables from the gauge parameter $\alpha$. Moreover, as previously discussed, BRST invariance also plays a key role in ensuring background field independence, i.e., the choice of the background field cannot change the value of observables. Thus, the present formulation is not consistent and a manifest $B$-gauge as well as BRST invariant construction of the action which eliminates infinitesimal Gribov copies in LCBG is mandatory. We work this out in the next section, following \cite{Dudal:2017jfw}.

\section{A non-perturbative BRST and $B$-gauge symmetric construction}
\label{non-p_BRST}

In this section we propose to follow the steps of \cite{Capri:2015ixa,Capri:2016aqq,Dudal:2017jfw} in order to construct a BRST-invariant action such that its path integral is restricted to a region where the Faddeev-Popov operator \eqref{fpop} is free of zero-modes. This is achieved by making use of the gauge-invariant field $a^{h}_{\mu}$ in the construction of the Horizon function. The field $a^{h}_{\mu}$ is constructed by the minimization of the functional
\begin{align}
f_{A}[u] = \Tr \int_{x} A^{u}_{\mu}A^{u}_{\mu}
\,,
\label{dim2cond}
\end{align}
over the gauge orbit. In \eqref{dim2cond}, $A^u_\mu$ denotes the finite gauge transformation of $A_\mu$ carried out by a group element $u\in SU(N)$. That is, the proposal is to select gauge field configurations that belong to a particular gauge orbit of $A_{\mu} = a_{\mu} + \bar{A}_{\mu}$ so that $f_{A}[u]$ is a local minimum. This can be carried out by expanding the gauge transformed configuration $A^{u}_{\mu}$, 
\begin{equation}
{A}^{u}_{\mu} = u^{\dagger} \left(a_{\mu} + \bar{A}_{\mu}\right) u + \frac{i}{g} u^{\dagger} \p_{\mu} u
\,,
\label{gtransfbfm0}
\end{equation}
in powers of $\theta$, with $u = \e^{ig\theta}$. Once again, in order to simplify the notation, we will only refer to the total gauge field $A_{\mu}$, but making explicit the fluctuation + background components whenever they need to be emphasized.

\subsection{Minimizing the dimension two operator}
\label{dim2gaugeinv}

Working with the total gauge field $A_{\mu}$ allows us to reproduce the results of \cite{Capri:2015ixa,Capri:2016aqq,Capri:2016gut} in the minimization of \eqref{dim2cond}. In the end, one should keep in mind that $A_{\mu} = a_{\mu} + \bar{A}_{\mu}$ in order to make connections with the BFM.

The functional \eqref{dim2cond} is gauge invariant for the field configuration $A^{h}_{\mu}$ with $h = \e^{ig\xi} = \e^{ig\xi^{a}T^{a}}$ that minimizes $f_{A}[u]$ along the gauge orbit of $A_\mu$. That is,
\begin{align}
f_{A}[h] := \min_{u} \Tr \int_{x} A^{u}_{\mu}A^{u}_{\mu} = 
\Tr \int_{x} A^{h}_{\mu}A^{h}_{\mu}
\,,
\label{dim2cond1}
\end{align}
is gauge invariant. 

The task of selecting $A^{h}_{\mu}$ for a gauge orbit so that $f_{A}[u]$ is at its \emph{global} minimum has not yet been accomplished and it is closely related to the existence of Gribov copies. The set of gauge field configurations that globally minimize \eqref{dim2cond} define the Fundamental Modular Region, which is completely free of Gribov copies. Yet one can impose the conditions to minimize $f_{A}[u]$ for a given gauge orbit perturbatively, \emph{cf.} \cite{Capri:2015ixa,Capri:2016aqq,Capri:2016gut} for details. It turns out that the conditions that must be satisfied to search for local minima of \eqref{dim2cond} are compatible with the conditions to eliminate infinitesimal Gribov copies. 

Let us consider a gauge transformation given by $v = he^{ig\theta}$, which has to be understood as power series of $\theta$. Formally, the gauge transformed field $A^{v}_{\mu}$ reads,
\begin{equation}
	A^{v}_{\mu} = v^{\dagger} A_{\mu} v + \frac{i}{g} v^{\dagger} \p_{\mu} v
	\,.
	\label{gtransfbfm}
\end{equation}
The explicit expression of $A^{v}_{\mu}$ up to terms of order $\mathcal{O}(\theta^{2})$ reads,
\begin{align}
	A^{v}_{\mu} &=
	A^{h}_{\mu}
	+ ig [\theta, A^{h}_{\mu}]
	+ \p_{\mu}\theta
	-\frac12 g^{2}  [\theta,[\theta, A^{h}_{\mu}]]
	\nonumber \\
	&
	+\frac12 ig [\theta,\p_{\mu}\theta] + \mathcal{O}(\theta^{3})
	\,.
\end{align}

Hence, the expression of $f_{A}[v]$ can be written as
\begin{align}
	f_{A}[v] &= 
	f_{A}[h]
	-2\Tr \int_{x} \theta\p_{\mu}A^{h}_{\mu}
	-
	\nonumber \\
	&
	- \tr \int_{x} \theta \p_{\mu} D^{h}_{\mu} \theta
	+ ig\tr \int_{x} \theta\p_{\mu}A^{h}_{\mu}\theta
	+{\cal O}(\theta^{3})
	\,,
\end{align}
where $\p_{\mu}D^{h}_{\mu}\theta = \p_{\mu}(\p_{\mu}\theta - ig[A^{h}_{\mu},\theta])$.

Now, the conditions for a local minimum of $f_{A}[v]$, are
\begin{align}
	\frac{\delta f_{A}[v]}{\delta \theta} \Bigg\vert_{\theta = 0} = 0 
	\quad \text{and} \quad
	\frac{\delta^{2} f_{A}[v]}{\delta \theta^{2}} \Bigg\vert_{\theta = 0} > 0 
	\,,
	\label{minimiscond}
\end{align}
which lead us to the well-known conditions
\begin{align}
	&
	\p_{\mu}A^{h}_{\mu} = 0
	\quad \text{and} \quad
	-\p_{\mu} D^{h}_{\mu}
	= {{\mathcal{M}}^{h} > 0}
	\label{Cop1}
	\,,
\end{align}
with $D^h_\mu \equiv \partial_\mu - ig [A^h_\mu,\,\,]$. That is, the total gauge configuration $A^{h}_{\mu} = a^{h}_{\mu} + \bar{A}^{h}_{\mu}$ must be transverse and the operator {${\mathcal{M}}^{h}$} positive definite. It must be clear that $A^{h}_{\mu}$ is \emph{not} the same as $A^{T}_{\mu}$, although both of them are transverse. In particular, $A^{T}_{\mu}$ is not gauge invariant. However, they coincide at zeroth order in the coupling constant {$g$ as discussed in the next subsection}.

\subsection{The dressed gauge invariant field}
\label{transnlocalahsec}

Here, the transversality condition in \eqref{Cop1} is used to derive the expression of $A^{h}_{\mu}$ as a non-local power series of $A_{\mu}$, and from now on the $A^{h}_{\mu}$ will be called the total dressed gauge field. Starting from
\begin{align}
A^{h}_{\mu} = h^{\dagger} A_{\mu} h + \frac{i}{g} h^{\dagger} \p_{\mu} h
\,,
\label{gtransfah}
\end{align}
{and, expanding} \eqref{gtransfah} up to quadratic powers of $\xi$, leads us to
\begin{align}
A^{h}_{\mu} &=
A_{\mu}
- \p_{\mu}\xi + ig [A_{\mu},\xi] + \frac12 ig [\xi,\p_{\mu}\xi] 
+\frac12 g^{2} [\xi,[A_{\mu},\xi]] 
\nonumber\\
&
+ \mathcal{O}(\xi^{3})
\,.
\label{ah01}
\end{align}
Imposing $\p_{\mu} A^{h}_{\mu} = 0$ we have
\begin{align}
\p^{2} \xi &=
\p_{\mu} A_{\mu} 
+ ig \p_{\mu} \left[ A_{\mu}, \xi \right]
- \frac12 ig \left[ \p^{2} \xi, \xi \right]
-
\nonumber \\
&-
\frac12 g^{2} \p_{\mu} \left[ [A_{\mu}, \xi], \xi \right]
+ {\cal O}(\xi^{3})
\,.
\label{xirdifeq}
\end{align}
Assuming $\xi$ is a smooth function of the coupling constant, which means that
$\xi$ can be written as $\xi = \sum_{n=0}^{\infty} g^{n}\xi_{n}$, one is
able to solve \eqref{xirdifeq} recursively in terms of $A_{\mu}$. That is, order by order in $g$ we
get 
\begin{align}
\xi_{0} &= 
\frac{\p_{\nu}}{\p^{2}} A_{\nu}  
\,;
\label{xi20}
\\
\xi_{1} &= 
\frac{\p_{\nu}}{\p^{2}}
i\left[ A_{\nu}, \frac{\p_{\sigma}}{\p^{2}} A_{\sigma} \right] 
\,;
\label{xi21}
\\
\xi_{2} &= 
{\cal O}(A_{\mu}^{3})
\,;
\label{xi2}
\\
&\hspace{2mm}\vdots
\nonumber 
\end{align}
For the sake of simplicity, we are keeping terms that are at most quadratic in the total field $A_{\mu}$. Thus, $\xi$ can be written as a power series of $A_{\mu}$,
\begin{align}
\xi &= 
\frac{\p_{\nu}}{\p^{2}}
\Bigg\{ 
A_{\nu}
+ ig \left[ A_{\nu}, \frac{\p_{\sigma}}{\p^{2}} A_{\sigma} \right]
+ \frac{1}{2}ig \left[ \frac{\p_{\rho}}{\p^{2}} A_{\rho}, \frac{\p_{\nu}\p_{\sigma}}{\p^{2}}A_{\sigma} \right]
\Bigg\}
\nonumber\\
&
+ {\cal O}(A_{\mu}^{3})
\,.
\label{xi01}
\end{align}

Finally, with \eqref{ah01} and \eqref{xi01} $A^{h}_{\mu}$ can by written as 
\begin{align}
	A^{h}_{\mu} &=
	\left( \delta_{\mu\nu} - \frac{\p_{\mu}\p_{\nu}}{\p^{2}} \right)
	\bigg\{
	A_{\nu} 
	+ ig \left[ A_{\nu}, \frac{\p_{\sigma}}{\p^{2}} a_{\sigma} \right] +
	\nonumber \\
	&
	+ \frac12 ig
	\left[
	\frac{\p_{\sigma}}{\p^{2}} a_{\sigma},
	\frac{\p_{\nu}\p_{\rho}}{\p^{2}} a_{\rho}
	\right] 
	+ {\cal O}(a_{\mu}^{3})
	\bigg\}
	\,.
	\label{ah03}
\end{align}
By construction, $A^{h}_{\mu}$ is gauge invariant and, as it will be seen in the next section, it is also invariant under $B$-gauge transformations. In other words, $A^{h}_{\mu}$ is invariant by shift transformations.

\subsection{The non-perturbative BRST symmetric action}
\label{npbrstactsubsec}

As we have just seen, by minimizing $f_{A}[v]$ {along the} gauge orbit of a particular configuration $A_{\mu}$, one gets that $A^{h}_{\mu}$ is {transverse}, and that {${\mathcal{M}}^{h} > 0$}. The gauge invariance of $A^{h}_{\mu}$ suggests that it is the ideal candidate to provide a BRST-invariant construction of the Horizon function. This was pursued in \cite{Capri:2015ixa,Capri:2016aqq,Capri:2016gut} and requires the definition of a region ${\Omega}^{h}$ which plays the analogue role of ${\Omega}^{T}$. It is defined by
\begin{equation}
{{\Omega}^{h}} = 
\left\{
{A_{\mu}\,,\bar{D}_\mu a_\mu = \alpha b} \,\,\vert\,\, \p_{\mu} A^{h}_{\mu} = 0, ~ {{\mathcal{M}}^{h} > 0}
\right\}
\,.
\label{greg02}
\end{equation}
Besides, since $A^{h}_{\mu}$ is {transverse} by construction, the Faddeev-Popov operator \eqref{fpop} is free of a large set of zero-modes once the path integral is restricted to ${\Omega}^{h}$, as was verified in {Sect.~\ref{0modessec}} for the region ${\Omega}^{T}$. 

Thus, proceeding with the restriction of the path integral to the region \eqref{greg02}, one gets\footnote{We omit the functional measures for the Faddeev-Popov ghosts and Lautrup-Nakanishi fields.}
\begin{align}
	Z_{{\Omega}^{h}} = \int [\mathrm{d}a]\, \e^{-{S}^{h}}
	\,,
	\label{hpathint}
\end{align}
with
\begin{align}
{S}^{h} &= {S}_{\FP} + {S}_{{H}^{h}} - \gamma^{4}d(N^{2}-1)
\,.
\label{nlgzaction1}
\end{align}
The Faddeev-Popov action is the same as the one given in the equation \eqref{fpaction0}, but here the action ${S}_{{H}^{h}}$ is different from the one in \eqref{nlhf}, with $A^{T}_{\mu}$ {being} replaced by $A^{h}_{\mu}$:
\begin{align}
{S}_{{H^{h}}} &= 
\gamma^{4}g^{2}
\int_{x} 
f^{abc}
A^{h,a}_{\mu} 
\left[ \left({\mathcal{M}}^{h}\right)^{-1}  \right]^{ce}
f^{dbe}
A^{h,d}_{\mu}
\nonumber \\
&=
\gamma^{4} {{H}^{h}(A)}
\,.
\label{nlhf01}
\end{align}
As in the case of ${S}_{{H}^{T}}$, the action \eqref{nlhf01} has two sources of non-locality: the operator $\left({\mathcal{M}}^{h}\right)^{-1}$, and the gauge invariant field $A^{h}_{\mu}$	{itself, see} \eqref{ah03}.

The localization of $\left({\mathcal{M}}^{h}\right)^{-1}$ {is achieved} by introducing a set of auxiliary fields $\{\bar{\varphi}, \varphi,\bar{\omega},\omega\}$ {as discussed in Sect.~\ref{grestrcsec},}
\begin{align}
\gamma^{4}\bar{H}^{h}_{\text{loc}} &=
\int_{x} 
\Big\{
\bar{\omega}^{a}_{i} 
\left({\mathcal{M}}^{h} \right)^{ab}
\omega^{b}_{i}
-\bar{\varphi}^{a}_{i} 
\left( {\mathcal{M}}^{h}\right)^{ab}
\varphi^{b}_{i}
\Big\}
+
\nonumber \\
&
+
\gamma^{2}g f^{abc}
\int_{x}
A^{h,a}_{\mu}(\bar{\varphi}^{bc}_{\mu} + \varphi^{bc}_{\mu})
\,.
\label{lhf02}
\end{align}
The second source of non-locality steams from $A^{h}_{\mu}$ and it can be made local by introducing a Stueckelberg-like field $\xi$ through $h = e^{ig\xi}$. The procedure is precisely the same as the used to get the expression \eqref{ah01}, provided the transversality condition $\p_{\mu}A^{h}_{\mu} = 0$ is ensured by a Lagrange multiplier,
\begin{align}
{S}_{\tau} = 
\int_{x} 
{\tau^{a}} \p_{\mu}(A^{h})^{a}_{\mu}
\,,
\label{ltfah0}
\end{align}
with the local field $(A^{h})^{a}_{\mu}$ being expressed as
\begin{equation}
{(A^{h})^{a}_{\mu}T^a =  h^{\dagger} A_{\mu} h + \frac{i}{g} h^{\dagger} \p_{\mu} h}
\,,
\label{ltfah0.1}
\end{equation}
and $h = e^{ig\xi}$. Thus, albeit $(A^{h})^{a}_{\mu}$ is a local composite operator, it is non-polynomial on the local fields $\xi$.

At last, a pair of anti-commuting fields {$\{\bar{\eta}^{a}, \eta^{b}\}$} must be introduced in order to consistently ensure that $\p_{\mu}A^{h}_{\mu} =0$\footnote{This pair of anti-commuting fields must be introduced in order to account for the Jacobian in the functional integration measure that appears to impose $\p_{\mu}A^{h}_{\mu}=0$, similar to the Faddeev-Popov gauge fixing process.},
\begin{align}
	{S}_{\bar{\eta}\eta} =
	- \int_{x}
	\bar{\eta}^{a} \left( {\mathcal{M}}^{h} \right)^{ab} \eta^{b}
	\,.
	\label{ahlocghst0}
\end{align}
Thus, the localized action reads,
\begin{align}
{S}^{h} &= {S}_{\FP} + \gamma^{4}{H}^{h}_{\text{loc}} + {S}_{\tau} + {S}_{\bar{\eta}\eta}
- \gamma^{4}d(N^{2}-1)
\,,
\label{lgzaction2}
\end{align}
where ${S}_{{H}^{h}}$ is the local one given by equation \eqref{lhf02}. Note that the transversality condition imposed by the Lagrange multiplier in equation \eqref{ltfah0} reflects a constraint on the Stueckelberg field, in the sense that once the $\xi$ field is eliminated by imposing $\p_{\mu}A^{h}_{\mu}=0$, the non-local expression \eqref{ah03} will be recovered. Further details about the localization procedure outlined above can be found in \cite{Capri:2017bfd}.

Now that path integral {domain} is properly restricted to $\bar{\Omega}^h$ by a suitable local action, one is able to verify that \eqref{lgzaction2} {is invariant} under the following BRST transformations,
\begin{align}
		s a_{\mu} &= - D_{\mu}c \,,
	\quad 
	s \bar{A}_{\mu} = 0 \,,
	\nonumber \\
	sc &= -\frac12 ig[c,c] \,,
	\nonumber \\
	s\bar{c} &= b \,,
	\quad
	sb = 0 \,,
	\nonumber \\
	s\bar{\omega}_{\mu} &=0 \,,
	\quad
	s\omega_{\mu} = 0 \,,
	\nonumber \\
	s\bar{\varphi}_{\mu} &=0 \,,
	\quad
	s\varphi_{\mu} =0 \,,
	\nonumber \\
	s \bar{\eta} &= 0 \,,
	\quad
	s\eta = 0 \,,
	\nonumber \\
	s\tau &= 0 
	\,,
	\label{nlbrsttrans}
\end{align}
and with
\begin{align}
	s\xi &= -c - \frac12 ig [c,\xi] + \frac{g^{2}}{12}[[c,\xi],\xi]
	+ \mathcal{O}(g^{3})
	\,.
	\label{nlbrsttrans1.1}
\end{align}
Notice that the invariance of \eqref{lgzaction2} under the BRST transformations \eqref{nlbrsttrans} {and \eqref{nlbrsttrans1.1}} steams, mainly, from the fact that $A^{h}_{\mu}$ {is} gauge invariant order by order in the coupling constant $g$ {which leads to $s(A^h)_\mu = 0$}. It must be clear {that} the gauge invariance of $A^{h}_{\mu}$ is also responsible for the invariance of \eqref{lgzaction2} under the $B$-gauge transformation. Besides, the auxiliary fields $\left\{ \bar{\omega}, \omega, \bar{\varphi}, \varphi, \bar{\eta}, \eta  \right\}$ and $\tau$, introduced to localize $\gamma^{4}{H}^{h}$, are BRST singlets (instead of doublets, as in the BRST-broken procedure).

\subsection{The $B$-gauge invariance of the action}
\label{bgaugesymmetricsec}

Alongside with the non-perturbative BRST transformation \eqref{nlbrsttrans}, the action \eqref{lgzaction2} is also invariant under a sort of non-perturbative $B$-gauge transformation, which is slightly different from the ones in \eqref{bgaugerule0} and \eqref{bgaugerule2}. Namely, 
\begin{align}
	\delta_{B} a_{\mu} &= -ig[a_{\mu},\theta] \,;
	\quad \text{and} \quad
	\delta_{B} \bar{A}_{\mu} = -\bar{D}_{\mu}\theta \,;
	\nonumber \\
	\delta_{B}c &= -ig[c,\theta] \,;
	\quad \text{and} \quad
	\delta_{B}\bar{c} = -ig[\bar{c},\theta] \,; 
	\nonumber \\
	\delta_{B} b &= -ig[b,\theta] \,;
	\nonumber \\
	\delta_{B} \bar{\varphi}_{\mu} &= -ig\bar{\varphi}^{mb}_{\mu} T^{b}[T^{m},\theta^{e}T^{e}]  -ig\bar{\varphi}^{am}_{\mu}T^{a}[T^{m},\theta^{e}T^{e}] \,;
	\nonumber \\
	\delta_{B}\varphi_{\mu} &=  -ig{\varphi}^{mb}_{\mu}T^{b}[T^{m},\theta^{e}T^{e}]  -ig{\varphi}^{am}_{\mu}T^{a}[T^{m},\theta^{e}T^{e}] \,;
	\nonumber \\
	\delta_{B} \bar{\omega}_{\mu} &= -ig\bar{\omega}_{\mu}^{mb}T^{b}[T^{m},\theta^{e}T^{e}]  -ig\bar{\omega}_{\mu}^{am}T^{a}[T^{m},\theta^{e}T^{e}] \,;
	\nonumber \\
	\delta_{B}\omega_{\mu} &= -ig{\omega}_{\mu}^{mb}T^{b}[T^{m},\theta^{e}T^{e}]  -ig{\omega}_{\mu}^{am}T^{a}[T^{m},\theta^{e}T^{e}]
	\nonumber \\
	\delta_{B}\bar{\eta} &= -ig[\bar{\eta},\theta] \,;
	\nonumber \\
	\delta_{B}\eta &= -ig[{\eta},\theta]
	\nonumber \\
	\delta_{B}\tau &= 0
	\,,
	\label{nlBgauge}
\end{align}
and
\begin{align}
	\delta_{B}\xi = -\theta -\frac12 ig [\theta, \xi] + \frac{g^{2}}{12} [[\theta,\xi],\xi] + \mathcal{O}(g^{3})
	\,,
	\label{nlBgaugexi}
\end{align}
which can be iteratively derived from the fact that $\delta_{B}A^{h}_{\mu} = 0$. The invariance of ${A}^{h}_{\mu}$ under $\delta_{B}$ can be understood from the fact that the $\delta_{B}$ variation of total gauge field is equivalent to a standard gauge transformation, while the total dressed gauge field $A^{h}_{\mu}$ is gauge invariant, then $\delta_{B}A^{h}_{\mu}$ must be null. Performing the $\delta_{B}$ variation of eq.\eqref{ah03} is equivalent, in fact, to perform a gauge transformation, whose invariance of $A^{h}_{\mu}$ is verified at \cite{Capri:2015ixa,Capri:2016aqq,Capri:2016gut}.

As can be seen, the transformation \eqref{nlBgauge}--\eqref{nlBgaugexi} differs from \eqref{bgaugerule0}--\eqref{bgaugerule2} by the transformation of the auxiliary fields $\eta$, $\bar{\eta}$ and the highly non-trivial $\delta_{B}\xi$.


It must be emphasized that any attempt to construct a Horizon function only in terms of the fluctuation gauge field $a^{h}_{\mu}$, instead of the total field $A^{h}_{\mu}$, must lead to the breaking of the shift symmetry.  

Thus, the action \eqref{lgzaction2} is local, BRST and $B$-gauge invariant, and effectively implements the restriction of the path integral to the region $\bar{\Omega}^h$ in the LCBG. With such properties, we can turn back to the question related to background-field independence. From eq.\eqref{lgzaction2}, it is clear that all terms but the gauge-gixing and Faddeev-Popov ghost terms involve the fields $\bar{A}_\mu$ and $a_ \mu$ under the combination $A_\mu \equiv \bar{A}_\mu + a_\mu$. Thus, all those terms are automatically invariant under the shift transformation \eqref{shiftsymmetry1}. As pointed out in Sect.~\ref{gpldw} and Appendix~\ref{Ap:ShiftWI}, the gauge-fixing together with the Faddeev-Popov ghost terms treat the background and fluctuation fields on an unequal footing rendering a non-trivial transformation under \eqref{shiftsymmetry1}. However, as in the standard gauge-fixed Yang-Mills theory, the resulting transformation is a BRST-exact term. Since \eqref{lgzaction2} enjoys BRST symmetry, this breaking is, again, harmless. Thus, the action \eqref{lgzaction2} provides a background-field independent results for observables. Conversely, one can see that if the Horizon function was not invariant under \eqref{shiftsymmetry1}, this would lead to a new source of shift-symmetry breaking which is not BRST exact and thereby this would spoil background-field independence. However, it is important to mention that although the present proposal ensures that shift-symmetry is automatically preserved by the terms that eliminate infinitesimal Gribov copies, it is fundamental to understand the renormalizability properties of the model. Such an analysis is beyond the scope of the present paper, but is essential in order to fully verify the consistency of background-field independence.

\section{Conclusions}

The present work is devoted to the study of the Gribov problem in Yang-Mills theories within the background field method in a linear covariant extension of the Landau-DeWitt gauge. Our central point is the status of the BRST and of the background gauge ($B$-gauge) symmetries of the action, as well as background-field independence of physical quantities, once the Gribov restriction is imposed. In principle we found out that both of these symmetries are broken by the Gribov restriction framework, being in agreement with the analysis in \cite{Dudal:2017jfw}. But then, we could verify that it is possible to address the Gribov problem consistently with a non-perturbative BRST and $B$-gauge transformations together with background-field independence of observables thanks to the sWI. This analysis complements the proposal in \cite{Dudal:2017jfw} and requires a careful study of the renormalization properties of the model.

Our first proposal was to decompose the total gauge field $A_{\mu} = a_{\mu} + \bar{A}_{\mu}$ into two components, $A_{\mu} = A^{T}_{\mu} + A^{L}_{\mu}$ being {transverse} and longitudinal according to \eqref{transgauge} and \eqref{longit}, respectively, and to impose the \emph{positive definite} condition to the operator ${\mathcal{M}}^{T}$, \eqref{Cop}. This operator is similar to the Faddeev-Popov operator \eqref{fpop} but with the total gauge field $A_{\mu}$ replaced by its {transverse} component $A^{T}_{\mu}$. In Section \ref{0modessec}, we could verify that by restricting the path integral to the region $\bar{\Omega}^{T}$, \eqref{greg01} (where ${\mathcal{M}}^{T} > 0$), the actual Faddeev-Popov operator is free of a large set of zero-modes.

The restricted path integral to $\bar{\Omega}^{T}$ is presented in Section \ref{grestrcsec}. Such a restriction is achieved by the introduction of the non-local Horizon function $\gamma^{4} \bar{H}^{T}(a^{T},\bar{A})$, \eqref{nlhf}, which could be localized with the introduction of the auxiliary fields $\{h^{a}_{\mu}, \xi^{a}_{\mu}\}$, $\{\lambda^{a}_{\mu}, \Lambda^{a}_{\mu}\}$, $\{v^{a}, \tau^{a}\}$ and $\left\{ \bar{\omega}^{a}_{i}, \omega^{a}_{i}, \bar{\varphi}^{a}_{i}, \varphi^{a}_{i} \right\}$, whose BRST transformation is given by the set of equations \eqref{brsta}--\eqref{brsttransf1}, while the $B$-gauge transformation is given by \eqref{bgaugerule0} and \eqref{bgaugerule2}. The consequence of restricting the path integral to $\bar{\Omega}^{T}$ is that both the BRST and the $B$-gauge symmetry are broken by the $\Delta_{\gamma^{2}}$ term (see equations \eqref{brstbreaking} and \eqref{bgaugegamma}). 

In the Section \eqref{non-p_BRST} we proposed a BRST and $B$-gauge invariant action restricted to the region where the Faddeev-Popov operator is free of a large set of zero-modes. The procedure was inspired by the works \cite{Capri:2015ixa,Capri:2016aqq}, where a particular configuration of the total gauge field that minimizes the {functional} \eqref{dim2cond}, called the dressed gauge field, was derived. As a consequence of minimizing $f_{A}[u]$, the dressed gauge field $A^{h}_{\mu}$ must be {transverse}, and {it} was found to be a non-local gauge invariant power series of $A_{\mu}$. A second consequence is that the operator ${\mathcal{M}}^{h} = -\p_{\mu}D^{h}_{\mu}$ must be positive definite, so that it defines a region ${\Omega}^h$, where the Faddeev-Popov operator is free of a class of zero-modes. In order to construct a BRST and $B$-gauge invariant action, whose path integral is restricted to ${\Omega}^h$, the standard Gribov restriction procedure must be imposed to the total dressed gauge field $A^{h}_{\mu}$, instead of only to the fluctuation configuration $a^{h}_{\mu}$. Since $A^{h}_{\mu}$ is gauge invariant by construction then we could verify, from equation \eqref{nlBgauge} to \eqref{nlBgaugexi}, that the local action ${S}^{h}$ is, indeed, invariant under background gauge transformation. This framework enabled us to address the issue of background-field dependence in the context where Gribov copies are taken into account. As discussed, BRST invariance plays a key role in the preservation of the shift-symmetry, which serves as an inspecting tool of background-field dependence.

The present work aimed at contributing to the very interesting and not yet fully understood question about the symmetry content of Yang-Mills theories within the BFM when the Gribov problem is taken into account (at least at the infinitesimal level). Thus, a detailed analysis of the renormalizability {properties} of the present model is left for a future work. Furthermore, it is also very interesting to investigate the effects of considering particular configurations of $\bar{A}_{\mu}$ as for finite temperature investigations. Finally, it is expected that the discussion presented in this work is related in a broader context. In particular, quantum-field theoretic formulations of quantum gravity, typically, have to face the issue of introducing a background metric. In the case of gravity, background independence is a founding principle and the analysis elaborated in the present work can be a stepping stone to tackle such an issue in quantum gravity, where a Gribov problem is expected to exist.

\section*{Acknowledgements}

The authors are grateful to David Dudal for useful discussions. This work was financed in part by the Coordenação de Aperfeiçoamento  de Pessoal de Nível Superior - Brasil (CAPES) – Finance Code 001. IFJ acknowledges CAPES for the financial
support under the project grant $88887.357904/2019-00$. ADP acknowledges CNPq under the grant PQ-2 (309781/2019-1), FAPERJ under the “Jovem Cientista do Nosso Estado” program (E26/202.800/2019), and NWO under the VENI Grant (VI.Veni.192.109) for financial support.

\appendix

\section{Derivation of the shift Ward identity \label{Ap:ShiftWI}}

In this appendix we explicitly derive the shift (or split) Ward identity \eqref{ShiftWI1} which plays a central role in ensuring, or better, controlling the background dependence in the theory. Observables should be completely blind to a particular choice of background. In the following, we derive the sWI and, next to that, we use it to prove the background independence of the free energy in pure Yang-Mills theories as a example of explicit application. 

Let us consider the partition function $\mathcal{Z}[J,\bar{A}_\mu]$ of Yang-Mills theories in the presence of a background gauge field $\bar{A}_\mu$ and external sources $J$ coupled to the elementary fields. Under \eqref{shiftsymmetry1}, it transforms as,
\begin{eqnarray}
    &&{\cal Z} [J,\bar{A}^\prime_\mu] = {\cal Z} [J,\bar{A}_\mu] - \int_x \epsilon^{a}_\mu (x) \frac{\delta {\cal Z} [J,\bar{A}_\mu]}{\delta \bar{A}^a_\mu (x)} \nonumber\\
    &=& {\cal Z} [J,\bar{A}_\mu] - \int_x \epsilon^{a}_\mu (x)\,{\cal Z}[J,\bar{A}]\, \frac{\delta {\cal W}[J,\bar{A}_\mu]}{\delta \bar{A}^a_\mu (x)}  \nonumber\\
    &=& {\cal Z} [J,\bar{A}_\mu] +{\cal Z}[J,\bar{A}] \int_x \epsilon^{a}_\mu (x)\, \frac{\delta \Gamma [J_\varphi,\bar{A}_\mu]}{\delta \bar{A}^a_\mu (x)}\,,
    \label{AP:Partition function01}
\end{eqnarray}
where we have used that ${\cal W} [J,\bar{A}] = {\rm ln}~{\cal Z}[J,\bar{A}]$ and,
\begin{equation}
    \Gamma [\varphi,\bar{A}] = -{\cal W} [J_\varphi , \bar{A}] + \sum_{\varphi}\int_x J_\varphi\cdot\varphi\,,
\end{equation}
which implies
\begin{equation}
\frac{\delta\Gamma[\varphi,\bar{A}]}{\delta \bar{A}^a_\mu} = - \Bigg(\frac{\delta {\cal W}[J_\varphi,\bar{A}]}{\delta \bar{A}^a_\mu}\Bigg)_{J_\varphi}\,.
\end{equation}
On the other hand, the partition function can be written as
\begin{equation}
\mathcal{Z}[J,\bar{A}^\prime_\mu] = \int [\mathrm{d}\mu_{\mathrm{FP}}][\mathrm{d}a^\prime]\mathrm{e}^{-S_{\rm FP}[\bar{A}^\prime,a^\prime;b,\bar{c},c]+\int_x J^a_\mu a^{\prime\,a}_\mu}\,,
\label{AP:PartitionFunction}
\end{equation}
with $[\mathrm{d}\mu_{\mathrm{FP}}]=[\mathrm{d}b][\mathrm{d}\bar{c}][\mathrm{d}c]$. Moreover,
\begin{eqnarray}
&& S_{\rm FP}[\bar{A}^\prime,a^\prime;b,\bar{c},c] 
= S_{\rm YM} [\bar{A}+a] + S_{\rm gf}[\bar{A},a, b] \nonumber\\
&+& S_{\gh}[\bar{A}, a, \bar{c},c]{+}\int_x\epsilon^a_\mu (x)\Bigg[ \frac{\delta (S_{\rm gf}+S_{\rm gh})}{\delta a^a_\mu}\nonumber\\
&-& \frac{\delta (S_{\rm gf}+S_{\rm gh})}{\delta \bar{A}^a_\mu}\Bigg]\,,
\label{AP:PartitionFunction2}
\end{eqnarray}
and
\begin{equation}
    \int_x J^a_\mu a^{\prime\, a}_\mu = \int_x J^a_\mu a^a_\mu + \int_x J^a_\mu \epsilon^a_\mu\,.
    \label{AP:PartitionFunction3}
\end{equation}
Expanding \eqref{AP:PartitionFunction} up to first order in $\epsilon^a_\mu$ leads to
\begin{eqnarray}
\mathcal{Z}[J,\bar{A}^\prime_\mu] &=& \mathcal{Z}[J,\bar{A}_\mu]\Bigg(1{-}\int_x\epsilon^a_\mu (x)\Bigg\langle \frac{\delta (S_{\rm gf}+S_{\rm gh})}{\delta a^a_\mu}\nonumber\\
&-& \frac{\delta (S_{\rm gf}+S_{\rm gh})}{\delta \bar{A}^a_\mu}\Bigg\rangle{+}\int_x\epsilon^a_\mu (x)\frac{\delta \Gamma}{\delta \hat{a}^a_\mu}\Bigg)\,.
\label{AP:PartitionFunction4}
\end{eqnarray}
Comparing \eqref{AP:Partition function01} with \eqref{AP:PartitionFunction4} yields
\begin{eqnarray}
\frac{\delta\Gamma}{\delta \hat{a}^a_\mu} - \frac{\delta \Gamma}{\delta\bar{A}^a_\mu} = \Bigg\langle \frac{\delta (S_{\rm gf}+S_{\rm gh})}{\delta a^a_\mu} - \frac{\delta (S_{\rm gf}+S_{\rm gh})}{\delta \bar{A}^a_\mu}\Bigg\rangle\,,
\label{AP:PartitionFunction5}
\end{eqnarray}
which can be written in a compact form by the introduction of the operator $\mathcal{B}^a_\mu (a , \bar{A})$ as
\begin{equation}
    \mathcal{B}^a_\mu (a,\bar{A})\circ := \frac{\delta}{\delta \hat{a}^a_\mu} - \frac{\delta}{\delta \bar{A}^a_\mu}\,,
    \label{AP:PartitionFunction6}
\end{equation}
where $\hat{a}$ stands for the expectation value of $a$ in the presence of the source $J$. Thus, eq.\eqref{AP:PartitionFunction5} is expressed as
\begin{equation}
    \mathcal{B}^a_\mu (\hat{a},\bar{A})\circ \Gamma = \langle \mathcal {B}^a_\mu (a,\bar{A})\circ (S_{\rm gf}+S_{\rm FP})\rangle\,.
    \label{AP:PartitionFunction7}
\end{equation}
In this work, we have adopted the following gauge fixing action (together with its corresponding Faddeev-Popov ghosts),
\begin{eqnarray}
 &&S_{\rm gf} + S_{\rm gh} = s\int_x \bar{c}^a\Big(\bar{D}^{ab}_\mu a^b_\mu - \frac{\alpha}{2}b^a\Big)\nonumber\\
 &=& \int_x\Big(b^a\bar{D}^{ab}_\mu a^b_\mu - \frac{\alpha}{2}b^a b^a + \bar{c}^a\bar{D}^{ab}_\mu D^{bc}_\mu c^b\Big)\,.
    \label{AP:PartitionFunction8}
\end{eqnarray}
This leads to
\begin{equation}
    \langle \mathcal{B}^a_\mu (a,\bar{A})\circ (S_{\rm gf}+S_{\rm gh})\rangle = - \Big\langle s\Big(D^{ab}_\mu\bar{c}^b\Big)\Big\rangle_{\bar{A},J}\,.
    \label{AP:PartitionFunction9}
\end{equation}
At vanishing source $J$, since the gauge-fixed Yang-Mills action is BRST-symmetric, the BRST-exact correlator in \eqref{AP:PartitionFunction9} vanishes. Together with \eqref{AP:PartitionFunction5}, this establishes the background-field independence of the free energy.


\bibliographystyle{utphys2}
\bibliography{refs}

\end{document}